# Uncertainties and Ambiguities in Percentiles and how to Avoid Them

Michael Schreiber

*Institute of Physics, Chemnitz University of Technology, 09107 Chemnitz, Germany. E-mail: schreiber@physik.tu-chemnitz.de*

The recently proposed fractional scoring scheme is used to attribute publications to percentile rank classes. It is shown that in this way uncertainties and ambiguities in the evaluation of percentile ranks do not occur. Using the fractional scoring the total score of all papers exactly reproduces the theoretical value.

**Introduction**

The recently proposed fractional scoring method (Schreiber, 2012) has been discounted by Leydesdorff (2012) as computationally intensive and as unnecessary complex. While his arguments are correct, I do not find his solution of attributing papers to percentile rank (PR) classes convincing. A Leydesdorff admits, his approach leads to an uncertainty depending of the applied counting rule. Although this uncertainty decreases with the number $N$ of documents as $1/N$, I find this uncertainty unsatisfactory.

For the linear problem of assigning quantiles, which are continuous variables, the middle of the uncertainty interval can be utilized. Leydesdorff (2012) proposes the same solution for the case of 100 PR classes, which he considers a linear problem, too. Strictly speaking, this is not so, but we rather have to deal with a step function comprising 100 steps. The fact that these steps are equidistant does not make the relation linear. In any case the decision of utilizing the middle of the uncertainty interval deals with the uncertainty also in this situation and should lead to a definite result. However, there are some cases where the result is ambiguous, namely at the steps, i.e., when the result equals the border between two PR classes, and therefore another decision has to be taken to arrive at a finally definite classification of the document in question. The solution is more complex in the clearly nonlinear situation of 6 PR classes for the bottom 50%, 50%-75%, 75%-90%, 90%-95%, 95%-99%, and top 1% publications, contributing with weights 1, 2, …, 6 to the $R(6)$ and $I3(6)$ indicators, proposed by Leydesdorff and Bornmann (2011).

**Problematic Examples**

In order to illustrate the uncertainty and the ambiguity let us consider a very simple example of only 5 documents and two PR classes, namely the bottom-50% class and top-50% class, generically attributing weights 0 and 1 to them. The example is visualized in Figure 1. Now let us investigate the $3^{rd}$ document. The lowest possible quantile with which this document can be described is 2/5 = 40%, e.g. counting all (namely 2) documents with worse performance as proposed by Leydesdorff and Bornmann (2011). Including the paper in question into the counting as suggested by Rousseau (2012) leads to 3 out of 5 documents determining the highest possible quantile of 60%. So there is the uncertainty between 40% and 60% as indicated by the yellow bar in the fourth row of Figure 1. Following Leydesdorff (2012), i.e. utilizing the middle of this interval in order to attribute the paper to a quantile leads to 50% and thus to an ambiguity if one aims at assigning one of the two PR classes, because it is a priori not clear whether the 50% quantile belongs to the bottom 50% or to the top 50% which are indicated in the first row of Figure 1. So here another definition is necessary in order to fix this ambiguity. Depending on this regulation for the third paper, either 2 or 3 papers will thus be assigned to the bottom-50% class, which means that either 40% or 60% fall into this class. Beside the uncertainty and the ambiguity, this deviation from the exact share of 50% is also unsatisfactory.

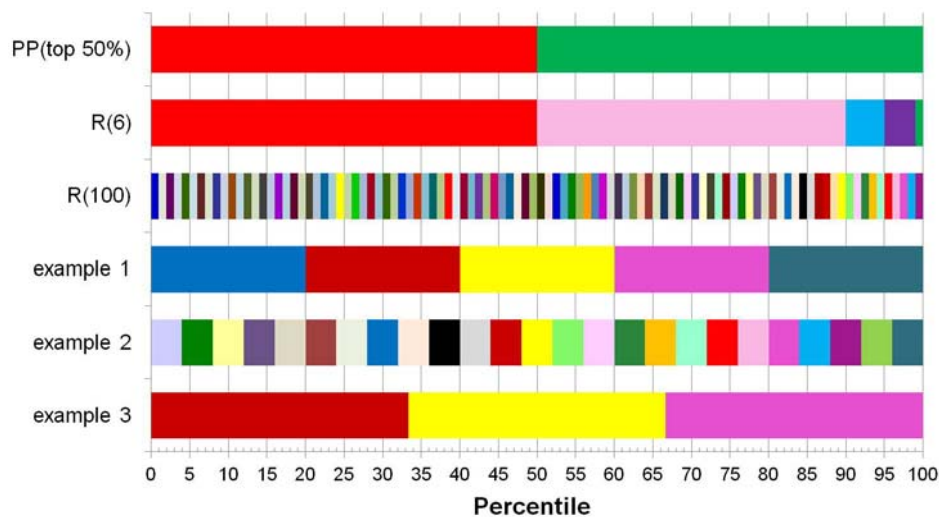

Figure 1. Percentile intervals for the PP$_{top\ 50\%}$ indicator (first row) and for the $R$(6) and $R$(100) indicators of Leydesdorff et al. (2011) (second and third row) and publication intervals for the 5, 25, and 3 publications in the first three examples discussed in the text. Vertical lines indicate every fifth boundary of the standard 100 percentile intervals.

A similar problem occurs for the 13$^{th}$ of 25 papers, with an uncertainty interval from 48% to 52% (see the fifth row of Figure 1), again leading to the same ambiguity of the 50% quantile. Considering the second out of 3 documents (as indicated in the last row of Figure 1), the uncertainty interval ranges from 33.33% to 66.67%, what translates to the 33rd percentile and the 66$^{th}$ percentile (compare the third row of Figure 1) if the usual rounding to the next lower integer is applied. The middle between 33 and 66, namely 49.5 would then be rounded to 49. Other ways of rounding or deciding on the percentile would possibly yield 33 and 67, with an ambiguous middle of 50, or 34 and 67 with a middle of 50.5 to be rounded to 51.

All described difficulties occur in the same way for the $R(6)$ indicator, visualized in the second row of Figure 1, because the border between the first two PR classes lies at the same position, namely at 50%.

In order to show that these are not only problems of a small total number of documents, I mention the 51st out of 101 documents with an uncertainty interval from 49.5% to 50.5% and the 500$^{th}$ paper out of 999 with an uncertainty between 49.95% and 50.05%. Again, this problem does not only occur for the attribution of the papers to the 2 PR classes, but in the same way with the same examples for the 6 PR classes as well as for the 100 PR classes. And it is not only a problem at the 50% quantile. For example at the 99% quantile one can find a similar uncertainty for the attribution of the 149$^{th}$ out of 150 papers to one of the 6 PR or one of the 100 PR classes, with an uncertainty interval from 98.67% to 99.33%.

**Solution**

All these problems occur, if one attempts to assign a certain quantile or percentile to the individual document. All these problems can be avoided, if one introduces a fractional attribution as proposed by Schreiber (2012). In this approach which is described in more detail by Waltman & Schreiber (2012) the above described uncertainty interval is not regarded as an uncertainty but rather as the significant and characteristic range which is associated with each paper. Then the interval is fractionalized and the respective fractions are attributed to the different PR classes. Graphically, this corresponds to the overlap between the "uncertainty interval" with the intervals representing the different PR classes in Figure 1. To be specific, for the first example discussed above, namely the third of 5 documents, the range from 40% to 50% thus belongs to the bottom 50% and the range from 50% to 60% naturally belongs to the top 50%. No averaging is allowed. As a result two and a half papers are assigned to the bottom class and two and a half papers to the top class. Thus the ideal proportion is achieved. If one utilizes 100 PR classes, then in this example a fraction of 0.05 of the third paper is assigned to every PR class from the 41st to the 60$^{th}$ class. Again the ideal proportions are obtained. In the continuous case, the respective interval of quantiles would be filled with a certain density.

In order to calculate the desired indicator, the now usually non-integer number of papers in a certain PR class has to be multiplied with the corresponding weight. Aggregating all papers in all PR classes this fractional scoring automatically reproduces exactly the correct total value of the integrated impact indicator $I3$, like 0.5 $N$ in the above used 2 PR case (with weights 0 and 1), 1.91 $N$ in the 6 PR scheme, and 50.5 $N$ in the 100 PR approach (assuming a weight of $k$ for the $k$-th PR class). The example provided by Leydesdorff (2012), namely the top paper in a set of 8 (visualized in Figure 2) achieves a score of 4.28 in the 6 PR situation, as determined by Leydesdorff. However, he then rounds this result and attributes the paper to the fourth of the 6 PR classes. Thus the full paper is assigned to this class with an integer weight of 4. This is completely different from the intention of fractional scoring. In the proposed fractional scoring scheme this paper is shared by 4 PR classes, compare the overlap of the intervals in the second and fourth row of Figure 2. As a result, the paper contributes a total score of 4.28 to the $I3(6)$ indicator and 4.28/8 = 0.535 to the $R(6)$ indicator. Thus the fractional attribution of the item to different classes results in a fractional scoring.

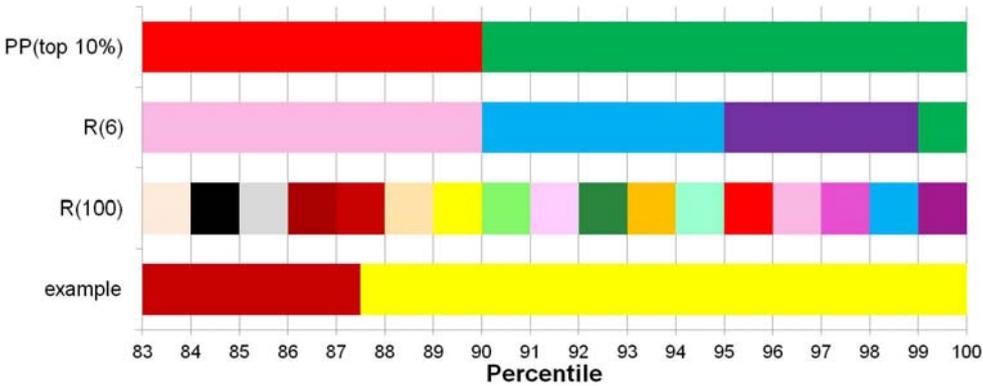

Figure 2. Percentile intervals for the PP$_{\text{top 10\%}}$ indicator (first row) and for the $R(6)$ and $R(100)$ indicators of Leydesdorff et al. (2011) (second and third row) and publication intervals for the 8 publications in the example discussed by Leydesdorff (2012). Vertical lines indicate the boundaries of the standard 100 percentile intervals. Note that only the upper end of the percentile scale is plotted, since the lower part is not interesting for the discussion. So in the second row only 4 of the 6 percentile intervals are shown, in the third row only 17 of the 100 percentile intervals appear, and in the fourth row only 2 of the 8 publications are indicated.

Utilizing 100 PR classes the top paper in a set of 8 covers the interval from 87.5% to 100%. Thus it fills one half of the 88th class and fully each of the 89th up to the 100th classes, compare the overlap of the intervals in the third and fourth row of Figure 2. The total score is $R(100) = 1178\% = 11.78$ equivalent to $I3(100) = 94.24$. Although the deviation is smaller, there is a difference from Leydesdorff's result which is based on the middle of the interval at 93.75 yielding a score of 93.75 if quantiles are utilized. Using 100 PR classes, this value would fall into the 94th class and thus should be given a weight of 94 (or, maybe, 93, if the rounding to the next lower integer is utilized).[1]

For the $PP_{top10\%}$ (percentage of top-10% publications) indicator discussed by Waltman et al. (2012) the paper count of the top paper in a set of 8 would be fractionalized, contributing with one fifth (2.5%) to the bottom-90% class and four fifth (10%) to the top 10%, as can be seen in Figure 2, comparing the first and fourth row.

**Discussion**

It is understandable that Leydesdorff (2012) has got the impression, that I mixed the normative and the analytical question, because in my previous publication I did not clearly separate between the fractional attribution of the papers to the PR classes and the subsequent attribution of weights. I hope that in the present communication I have clarified how fractional scoring works. I admit that the fractional attribution is computationally expensive. But the advantage is that it avoids the uncertainties and ambiguities. It always yields the correct total score. To the best of my knowledge, it is the only scheme that reproduces the theoretical total score perfectly.

In the present investigation I have used examples with very few papers in order to present the problems and the solution in a clear way. But I note that these problems are not academic questions with relevance only in the case of small numbers of publications. For larger datasets rather often many papers are tied regarding their citation counts. Waltman et al. (2012) discuss the application of the fractional scoring scheme for the $PP_{top10\%}$ indicator and show that it is especially important in the case of tied papers to attribute the papers fractionally to the respective PR classes. They show in particular that otherwise a comparison between different fields can be strongly distorted.

In conclusion, fractional scoring is strongly recommended.

---

[1] It remains unclear, why Leydesdorff (2012) rounds the score of 4.28 to 4 in the case of 6 PR classes, but does not round 93.75 in the case of 100 PR classes.

However, probably the main problem is that it requires a change of perspective. Papers are not attributed at certain quantile points in the range between 0 and 1, but attributed to intervals. Mathematically speaking, they are represented by a paper density over a respective interval instead by a δ-function. The overlap of this interval with the intervals corresponding to the various PR classes determines the fractions with which the paper belongs to these classes as discussed by Waltman and Schreiber (2012). In the fractional scoring approach, these fractions are weighted with the respective weights of the PR classes in order to determine their contribution to the indicator of interest. In conclusion this procedure means that the traditional point of view has to be given up. In my opinion, this is appropriate in view of the listed advantages.

Finally I describe an example from real life: Let us consider a glass of water or wine which is 50% filled. A frequent, though futile debate raises the question, whether this glass is half empty **or** whether it is half full. Using as above 2 PR classes for the bottom 50% and the top 50%, all the traditional scoring schemes would attribute the glass completely either to the bottom or to the top class, depending on the way, how the uncertainties and the ambiguity are treated. This means that the glass is considered either completely empty or completely full. Most of us, if not all of us will agree that this is nonsense. Applying the fractional scoring scheme, the glass will be considered to be 50% empty **and** 50% full, which I believe is the most reasonable result.